*Fuel tax loss in a world of electric mobility: A window of opportunity for congestion pricing*

Thi Ngoc Nguyen[*a], Felix Müsgens [a]

[a] *Brandenburgische Technische Universität Cottbus-Senftenberg, Siemens-Halske-Ring 13, 03046 Cottbus, Germany*

**Abstract**

The continued transition towards electric mobility will decrease energy tax revenues worldwide, which has substantial implications for government funds. At the same time, demand for transportation is ever increasing, which in turn increases congestion problems. Combining both challenges, this paper assesses the effectiveness of congestion pricing as a sustainable revenue stream to offset fuel tax loss in 2030 while simultaneously enhancing efficiency in the transport sector. A congestion-based toll that is road-and-time-variant is simulated for the greater Berlin area in Germany using the multi-agent transport simulation (MATSim) software. Through the simulation results, this paper quantifies the impacts of the toll on the governmental revenue, traffic management, environment, social welfare, and the distribution effects. We find that the revenue from congestion tolls in a metropolitan area can compensate the reduction in passenger car fuel tax. Furthermore, a remarkable welfare surplus is observed. The toll also successfully incentivises transport users to adjust their travel behaviour, which reduces traffic delay time by 28%. $CO_2$ emissions as a key metric for decarbonisation of the transport sector decrease by more than 5%. The analysis of the distribution effects suggests that a redistribution plan with a focus on the middle-low-income residents and the outer boroughs could help the policy gain more public acceptance.

[*] Corresponding author

*Email address:* nguyen@b-tu.de (Thi Ngoc Nguyen)

Correspondence to: Brandenburgische Technische Universität Cottbus-Senftenberg, Siemens-Halske-Ring 13, 03046 Cottbus, Germany.

Highlights

- Congestion tolls can offset EV-caused fuel tax shortfalls in metropolitan regions.
- The toll reduces traffic delay time by 28% and $CO_2$ emissions by over 5%.
- Toll revenue redistribution should focus on the middle-low-income residents.
- Investments on the transport systems should upgrade the outer boroughs.

1. Introduction

The rapid transition to electric vehicles (EVs) is reshaping the global transportation sector, with significant implications for society and the environment. While considerable literature has examined the environmental and technological impacts of EV adoption (Isik et al., 2021; Powell et al., 2022), the effects on government revenues, especially energy taxes, have been largely overlooked. This gap in the research is concerning and timely, given the escalating financial demands of climate change mitigation and adaptation, and the tightening constraints on government budgets worldwide.

Energy taxes are an important source of public finance to provide social services and public goods. These taxes are thus crucial for the national welfare. In Europe, these taxes make up more than three-quarters of all environmental taxes (Eurostat, 2023) and in many developing countries, they represent nearly 100% of such revenues (Matheson, 2021). However, global energy tax revenues have been declining. Between 2016 and 2020, the share of energy taxes out of global GDP dropped by nearly 10% (OECD, 2023). This reduction is primarily attributed to decreased fuel consumption in road transport (European Commission, Directorate-General for Taxation and Customs Union, 2022). Transport fuel taxes are the largest component within the energy tax structure, far surpassing other elements such as carbon taxes and electricity taxes (Matheson, 2021). The drop in fuel taxes is expected to accelerate as electric cars are replacing internal combustion engines (ICEs) around the world.

According to IEA (2024), sales of electric vehicles (EVs) are increasing globally with 40 million EVs on the road in 2023 (IEA, 2024). The annual average growth rate of the global EV stocks, previously estimated at 30–32% until 2030 (IEA, 2021), is expected to be surpassed and EVs could reach over 20% of the total car sales in 2024 (IEA, 2024). While this shift is crucial to reduce transportation sector emissions, it also poses a significant challenge to government revenues. Particularly in regions such as Europe, where both a high share of energy taxes out of total taxation and a major expansion of EV markets are observed, the revenue loss could have important impacts.

Our study highlights that the reduction in fuel taxes due to increased EV use will be only marginally offset by an increase in electricity taxes, creating a fiscal shortfall that necessitates innovative revenue solutions. In Germany, for example, our estimations show the tax rate per kilometre travelled is around €0.03/km for fuel taxes and less than €0.005/km (6 times less) for electricity taxes. In 2030, fuel taxes in this country could be reduced by around €12.5 billion compared to 2020 due to EV growths. Meanwhile, electricity taxes from EVs are projected to increase by only €0.75 billion above the 2020 level (Figure *1*). Similarly, in a case study of the USA (Jenn et al., 2015), an annual revenue shortfall of up to 900 million dollars (due to EV diffusion) was expected already until 2025.

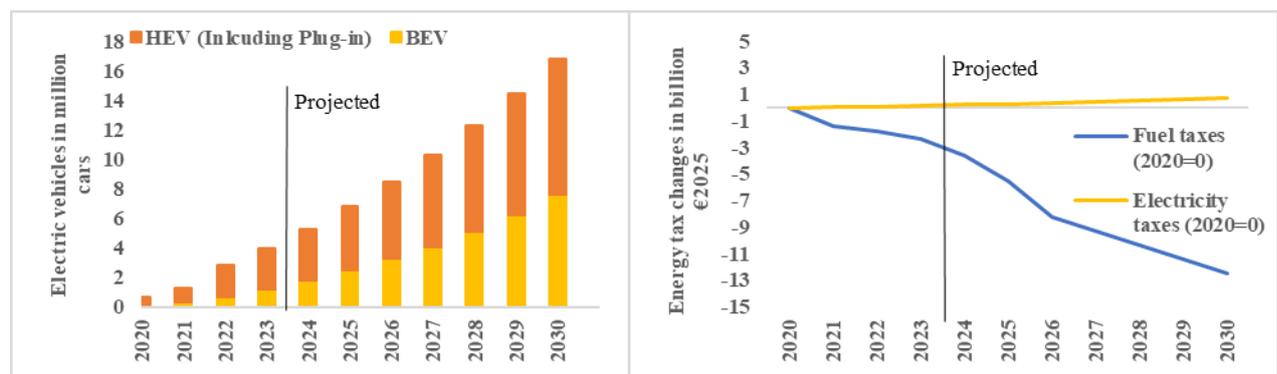

Figure 1: Electric vehicle diffusion (left panel) and the changes in energy tax from passenger cars (right panel) in Germany between 2020 and 2030. Data from 2024 to 2030 are projected. Source: Own illustration. BEV = Battery electric vehicle. HEV = Hybrid electric vehicle (including Plug-in). All monetary values are inflation adjusted based on CPI (2025 = 100). Data is estimated based on the scenario of a dynamic growth of EVs in Germany (Betz et al., 2021). Details about our estimation is presented in section 4.1.

The need for more focused research and policy discussions on sustainable revenue streams in the world of electric mobility is becoming increasingly urgent. Although there is no explicit mandate to offset declining revenues within the transportation sector, it is likely that a significant portion of this shortfall will need to be compensated within this domain. Taxation on EVs (Noll et al., 2024) and car registration fees or per mile surcharges (Jenn et al., 2015) have been suggested, but these would make EVs more expensive and contradict with the policies encouraging their adoption (Aasness and Odeck, 2015; Bjerkan et al., 2016). Increasing the gasoline tax rate was also discussed as a solution to fill in for the loss (Tscharaktschiew, 2015). However, this might incentivise an even stronger shift to EVs and reinforce the fuel tax revenue reduction.

Concurrently, traffic congestion remains a persistent global problem. Unlike emission, which would be significantly reduced by EVs, congestion remains an externality for both conventional and electric vehicles and was shown to have the leading responsibility for all transport related externalities (Borger et al., 1996; Kaddoura et al., 2017). Studies showed that compared to normal driving conditions, traffic congestion leads to 80% more energy consumed (Cramton et al., 2018) and more $CO_2$ emissions are observed (Bharadwaj et al., 2017).

Congestion pricing is widely recognised in the literature as an effective solution to enhance the transport system's efficiency (Palma and Lindsey, 2011). A well-designed congestion toll system would mitigate the traffic jam (Meng and Liu, 2012; Zheng et al., 2016), which leads to less energy waste and therefrom emissions (Yang et al., 2023) as well as noise pollution (Kaddoura, 2015; Cavallaro et al., 2018). It also leads to a more efficient use of the road system, thus reducing the needs for expensive road expansion and saving funds for other public goods and services. Many case studies around the world such as Singapore, London, Stockholm, Milan, and Gothenburg have shown the positive impacts of this policy (Litman, 2018). In the context of electric mobility, congestion pricing is among a few that raise additional revenues without impeding the growth of EVs. It has less adverse impacts on the EV adoption willingness than upfront cost related fees (Shafiei et al., 2018) and encounters less political resistance than increasing fuel taxes further (Yan, 2018).

Although congestion pricing offers significant benefits, it remains underutilized by policymakers, partly due to the public's uncertainties about its impacts on social welfare. Given the pressing need for alternative revenue streams beyond fuel taxes and the high efficiency potential of congestion pricing, our paper hypothesizes that there is a window of opportunity for congestion pricing to effectively address both mentioned issues simultaneously.

Consequently, our study examines the effectiveness of congestion pricing as a sustainable revenue stream to offset energy tax losses resulting from a transition to electric mobility while simultaneously enhancing efficiency in the transport sector. The "effectiveness" is measured through the following questions:

(i) To what extent can a congestion-based road toll, with no constraints on total toll revenue, offset fuel tax losses?

(ii) What are the impacts of the road toll on the traffic operation and the environment?

(iii) What are its impacts on social welfare and distribution effects, both vertically (i.e. between income groups) and horizontally (i.e. between different areas)?

To answer these questions quantitatively, we analyse a case study of the greater Berlin area in Germany in 2030, which includes the adult population residing in Berlin state and the commuters from Brandenburg state to Berlin. This region has a dynamic diffusion of EVs and the traffic congestion issue is also critical (Erdmenger et al., 2023). The year 2030 was selected as it is a key milestone set by various international agreements on climate change (e.g., Paris Agreement, The United Nations' 2030 Agenda for Sustainable Development, The European Green Deal), which strongly encourage the transition to electric mobility.

We begin by estimating the energy tax revenues (including both fuel and electricity taxes) from passenger cars in the region for 2030, assuming dynamic growth in EV adoption based on Betz et al. (2021). Next, we simulate the regional traffic conditions with and without a congestion toll using the Multi-Agent Transport Simulation (MATSim) software (Horni et al., 2016), which is particularly suited for simulating real-world case studies and capturing complex traffic dynamics. The simulation results provide congestion toll revenues, which we compare against the projected reduction in energy tax revenues in 2030 to assess whether they can compensate for the shortfall. Additionally, by comparing outputs across different scenarios, we examine the potential impacts of congestion pricing.

In our analysis, transport users are simulated as individual agents, each with distinct travel plans and scoring functions. This allows us to account for individual heterogeneity—a critical factor when examining the distributional effects of policies like congestion pricing. All agents interact with one another and respond to the tolls within a simulated physical environment. Through an evolutionary iterative process, each agent optimizes their travel decisions to maximize their personal utility. By analysing how agents adjust their behaviour in response to tolls, we gain a detailed and nuanced understanding of how congestion pricing might influence traffic patterns, user choices, and the social welfare.

We explore three scenarios. First, a "Reference" scenario replicates the work of Ziemke et al. (2019), simulating 10% of the adult population in the region travelling based on a survey of their real life daily travel plans. There is no congestion toll in this scenario. Second, a "Congestion" scenario charges car drivers depending on the traffic delay situation on the road at the time. Everything else, including the transport infrastructure and public transport costs, is identical with Reference scenario. Third, in a Congestion+ scenario, the road capacity is expanded by 10% and public transport costs are reduced by 20% compared to the other two scenarios. Other assumptions are equal to the "Congestion" scenario. The "Congestion+" scenario analyses a situation where traffic delays are partially tackled by investment in the road infrastructure and subsidies for public transport, which could lead to less revenue from congestion pricing.

This study analyses both vertical and horizontal distribution effects. Vertical distribution implies the redistribution of tax revenue between income groups. Congestion pricing can be regressive since people with high income often have a higher value of time and gain more from less travel delay (Small, 1983). At the same time, the more affluent people could be charged more and the tax revenues could be invested in improving the public transport system, thus benefiting the low-income group (Santos and Rojey, 2004) and making the policy progressive. Horizontal distribution effects refer to the impacts of the policy within the same income group (Eliasson and Mattsson, 2006). For example, people who live further from the work place (usually the city centre) have to travel more and are thus more likely to pay higher congestion fees than those living around the city centre (He et al., 2021). The former group is therefore more presumably to vote against the policy, even though the two groups can be in the same income interval. Therefore, measuring horizontal in addition to vertical distribution effects helps identifying the winners and losers from the policy more accurately. This is crucial to assess the potential of introducing the policy, and also to formulate the revenue redistribution plan and enhance public acceptance.

"Interval-based list pricing (LP)" is the toll charging approach that will be simulated. The toll is road-time-variant and congestion-based. This approach has been shown to reduce traffic delay more efficiently than flat rate fee or distance-based pricing (Simoni et al., 2019) but stay simple and convenient to compute (Kaddoura and Nagel, 2019). This balance between efficiency and complexity is important to the public understanding and acceptance of the policy (Tirachini and Hensher, 2012). It also reduces the implementation and monitoring costs, making the implementation of the policy politically and economically feasible. In this study, the toll will be introduced for passenger cars only, which has the main responsibility for the traffic congestion. This mode also dominates all the others in both EV adoption and fuel tax contribution and should be the focus of the analysis.

This paper has several novel contributions:

- First, this is the first study to analyse congestion pricing as a viable alternative to fuel tax in the context of a rapid transition to electric mobility. We provide a comprehensive assessment of the policy's suitability and feasibility. In addition to the contribution to government revenues, important impacts of congestion pricing regarding the traffic management, environmental benefits, social welfare, and distribution effects are quantified. In particular, not only vertical distribution effects between income groups are analysed, but horizontal distribution effects between different spatial zones are also studied. Based on this, the toll revenue redistribution scheme can be properly formulated to enhance the public acceptance.

- Second, through the analysis, our paper demonstrates a "window of opportunity" for intelligent road pricing: On the one hand, congestion pricing is among a few that raise additional revenues without impeding the growth of EVs. On the other hand, the advantages of intelligent road pricing for a more efficient transportation system are proven, which might be increased even further thanks to technological development. If rightly communicated, such a policy could be largely accepted by the public. Given that congestion pricing has been long discussed but remains underutilized, the emerging shortfall in energy tax revenue presents a timely opportunity to reintroduce this policy for serious consideration and potential implementation.

- Finally, the findings and methodologies in this study have global relevance. The transition to EVs is a worldwide phenomenon that will undoubtedly reduce global fuel tax revenues – a crucial source of revenues for most countries. When intelligent road tolls such as congestion pricing could effectively solve this problem for the region studied, they hold considerable promise for other regions as well. Furthermore, the policy assessment method applied in this study can also be replicated for other case studies and is thus an important tool for politicians, local authorities, researchers, and other stakeholders.

The structure of the rest of the paper is as follows. Section 2 discusses the literature on congestion pricing methods. Section 3 provides an overview of the case study and simulation setup. Section 4 explains the key methodologies applied in the paper. Section 5 discusses the simulation results. Section 6 concludes the paper.

**2. Literature review on congestion pricing**

Road tolls have been widely suggested as an efficient solution to reduce traffic congestion as well as other transport related externalities (Palma and Lindsey, 2011). There are many strategies for congestion pricing that are currently applied worldwide. Most of them could be classified as "flat toll", "distance-based toll", or "dynamic toll" (Simoni et al., 2019), as illustrated in Figure 2.

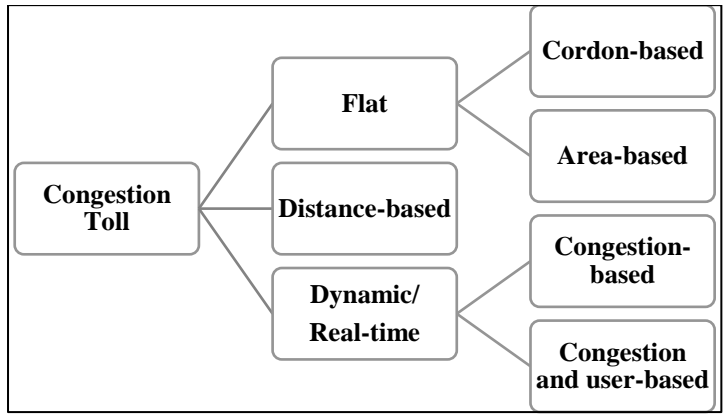

Figure 2: Popular congestion pricing strategies. Source: Own illustration.

A flat toll is a constant fee applied for an area during a specific time. This is called "cordon-based" when vehicles enter or cross the area, or "area-based" when vehicles drive inside the area. Such tolls have been adopted in many cities such as Singapore, London, Stockholm, Milan and Gothenburg (Lehe, 2019). Another approach is to charge the fee based on the distance travelled by vehicles (Wangsness, 2018). Distance-based tolls are often suggested as a solution to raise revenue, especially when the fuel tax is reduced because more fuel-efficient and electric vehicles enter the roads (Simoni et al., 2019). However, this policy could worsen the congestion issue if drivers are incentivised to take shorter (but often with more congestion) routes (Liu et al., 2014).

Both flat tolls and distance-based pricing neglect the dynamic traffic conditions, which are particularly important in optimising the congestion toll rate. Ideally, congestion pricing should be dynamic, reflecting real-time traffic conditions and road user characteristics (e.g., income, value of time, alternative modes) (Simoni et al., 2019). By incorporating these information into the toll-setting process, it is possible to design tolls that maximize social welfare. One approach to achieving welfare-maximizing tolls, which consider traffic conditions and allows for individualized fees, is marginal cost pricing (MCP). This toll is computed based on the queuing dynamics at the bottlenecks and equals the marginal external congestion costs caused by the agent depending on its position in the queue (Rafferty, P., Levinson, D., 2004). There have been several studies analysing this road or link-, time- and user-specific toll (Kaddoura, 2015; Chen et al., 2021). This method requires the information of all vehicle's position in the traffic network and the computation of the toll is complex. The implementation and monitoring of such a pricing system is costly and can meet public objection due to its complexity as well as the data protection issue (Erdmenger et al., 2023).

"Interval-based list pricing (LP)" is a congestion pricing method that balances well between efficiency and complexity. It adjusts the toll automatically based on the congestion level on the road. The toll is the same for all agents in the same road and time interval. Compared to MCP method, this toll is more convenient to compute but have equivalent effects on the social welfare (Kaddoura and Nagel, 2019). This congestion pricing approach will be simulated in our study.

There were several studies simulating LP congestion pricing (Kaddoura et al., 2017; Kaddoura and Nagel, 2019; Chen et al., 2021), measuring the toll revenue as well as the impacts on traffic delay. In Chen et al. (2021), three congestion toll strategies were simulated for a case study of Melbourne. The LP approach was compared with cordon tolls. The study showed that LP congestion pricing successfully incentivised transport users to reduce their travel inside the congested areas or shift their travel times, thus reducing traffic congestion. Kaddoura and Nagel (2019) analysed MCP and LP congestion pricing in the greater Berlin region. The work showed that the LP approach is superior to the MCP in both simplicity and social welfare achievement. The LP was also proved to react better to the dynamics of the traffic system and agents' learning. In Kaddoura et al. (2017), LP congestion toll was simulated for the Greater Munich area. This work examined the noise and air pollution pricing for the transport users. Different combinations of choice dimensions were experimented. It showed that an ideal pricing scheme should charge all externalities at the same time. Furthermore, it also highlighted the importance of analysing the distribution effects of the policy.

Previous studies have consistently concluded that congestion pricing effectively alleviates traffic congestion and generates welfare surplus. This paper contributes to the existing literature by examining the policy's impact on government revenue within the context of increasing electric mobility. In addition to the impacts of the policy on the society on a macro level, the effects on individuals (e.g., characterised by travel preferences, income groups, home locations) are also examined.

3. **Case study**

We simulate the daily traffic demand of adults (at least 18 years old) in the greater Berlin area in Germany, including the residents in Berlin state and the commuters from Brandenburg to Berlin. The greater Berlin area is the capital region of Germany and the third-largest metropolitan region of the country, where traffic congestion is a substantial issue (Erdmenger et al., 2023). Furthermore, this region also has a dynamic diffusion of EVs and has the highest number of EVs registered in Germany as of January 2024 (Statista, 2024). Recently, Berlin state announced their plan to enhance the charging capacity by almost sevenfold by 2030 (berlin.de, 2024). Discussing congestion pricing in this region in the context of rapid EV diffusion is particularly relevant and can provide useful knowledge to other regions worldwide.

Figure 3 shows the transport network of the region, which includes all major and minor roads and public transit lines. The most popular modes of transport are included in the simulation, i.e., car (using cars as drivers / passenger car), ride (using cars as paid services / taxis), bicycle, walking, and public transport (PT) (based on the actual schedule). The freight transport is also simulated as background traffic.

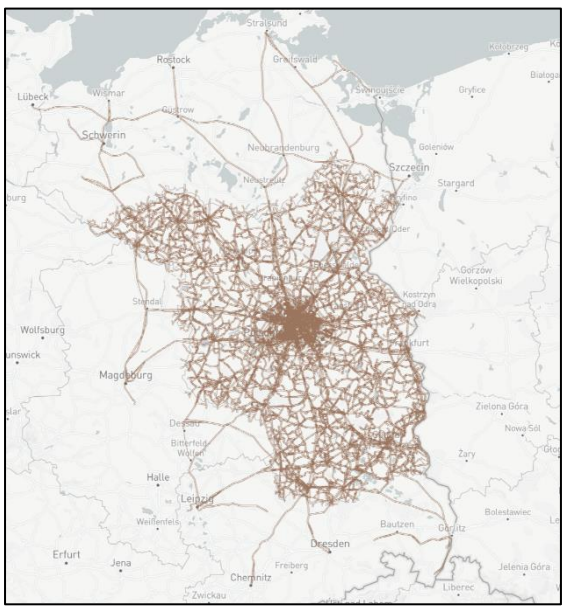

Figure 3: Traffic network in the greater Berlin area in Germany. Source: Own illustration.

Due to the cost of computation, only 10% of the population is simulated with the resulting travel behaviour assumed to be representative for the full population. The "Reference scenario" simulates the traffic demand without any tolls. This replicates the work of Ziemke et al. (2019), which was validated successfully against the real data. This validation ensures that the policy analysis based on the simulation results is realistic and applicable. The reference is simulated for 1000 iterations, which is deemed sufficient for transport users to optimise their travels (Agarwal and Kickhöfer, 2018).

A "Congestion scenario" introduces a toll controller optimizing traffic delay in the simulation. As transport users now have optimised their travels through the previous 1000 iterations, the simulation continues for only 500 more iterations. During the simulation, agents are informed from previous iterations about the average travel time of each road segment and toll fees. With this information, agents can adjust their travels to maximise their utility. For example, agents may take a longer but less congested route, depart earlier or later than normal, or change from car to other modes (e.g., bike, walk, PT) to avoid the congestion toll. The utility and travel behaviour of all agents are recorded and analysed in comparison with Reference case, which shows the impacts of introducing congestion pricing in the region.

Finally, in a "Congestion+ scenario", the same setting as in the Congestion scenario is used, except that the road capacity is increased by 10% and the daily constant cost of PT is reduced by 20%. This scenario examines the contribution of traffic congestion toll to the national revenue when there is an upgrade in transport infrastructure and a subsidy for PT that could lead to less congestion (and thus less toll revenue).

For all three scenarios, we assume that in a world of massive diffusion of EVs, the charging stations are as available as gas stations and the travel behaviours of EVs could resemble normal cars. Therefore, we conduct no separate simulation of EVs' charging behaviour. Note that the focus of the paper is on passenger cars. For the rest of the paper, unless stated otherwise, the term "car" will always indicate the passenger car.

## 4. Methods

This section presents our methodology. First, we explain our estimation of the energy taxes from passenger cars in Germany and the two states Berlin-Brandenburg[1]. Then, an overview of the simulation framework (MATSim) is given. Next, the formulation of the income-dependent utility function, which is used to score agents' travel plans and decide optimal travel strategies, is described. Together with this, the conversion between monetary and utility terms from the simulation output data are explained. After that, the LP congestion pricing method is illustrated. Finally, the upscale of the simulation output data is presented.

### 4.1. Estimation of energy taxes from passenger cars

---

[1] Note that the greater Berlin, which is the basis for the simulation of congestion toll revenues, encompasses all of Berlin but only parts of the traffic in Brandenburg. However, the specific energy tax revenue data for the greater Berlin area alone is not available to the best of our knowledge. Therefore, the data of the two states are used instead, making the estimated energy tax loss (for all of Berlin and Brandenburg) higher than that for the simulated area (the greater Berlin). Thus, if the congestion pricing revenue from the greater Berlin area can cover the energy tax loss for the two states, it obviously compensates that for the simulated area.

We estimate energy taxes from passenger cars, including fuel and electricity taxes, in the whole of Germany as well as for the two federal states Berlin and Brandenburg based on the guidelines on taxing energy use in Germany provided by OECD (OECD, 2019).

The taxes equal the multiplication of the resources consumed with the tax rate for each type of resource. Resources comprise conventional fuel (mainly diesel and gasoline) and electricity consumed for driving by cars. The resource consumption is estimated based on the data of car stocks by powertrain, mileage, and the average fuel or electricity consumption of each car type per kilometre. This is similar to the methodology applied in the work of Betz et al. (2021) to forecast the resource demand for Germany until 2035.

The data of car stocks by powertrain in Germany is taken from the work of Betz et al. (2021), which assumes a dynamic growth of EVs in Germany. According to this work, the car stock in 2030 in Germany comprises 63% internal combustion engine (ICEs, fuelled by petrol or diesel), 8% hybrid EVs (HEVs), 12% plug-in hybrid EVs (PHEVs) and 17% battery EVs (BEVs). We assume that identical changes will also occur in the simulated region. Based on Betz et al.'s estimations of the annual car stocks by powertrain in Germany, we compute the annual growth rate of each vehicle type until 2030. Using this information and the actual data of car stocks in Berlin-Brandenburg states in 2023, we estimate the stock of each vehicle type in the region until 2030. The car stock, the mileage, and the tax rate data are then used to calculate the fuel and electricity taxes as followed:

The fuel tax for each year is computed as the sum of fuel taxes from all the passenger cars that are fuelled by petrol or diesel, including the following vehicle types: ICE-petrol, ICE-diesel, HEV-petrol, HEV-diesel, PHEV:

$$fuel\_tax_y = \sum_i fuel\_tax_{i,y} \qquad (1)$$

with $i$ being the vehicle type and $y$ being the year.

For each vehicle type, fuel tax is computed as the multiplication of the tax rate per fuel litre (€/l) and the annual distance travelled or mileage (km) and the average fuel consumption per kilometre (l/km) as follow:

$$fuel\_tax_{i,y} = fuel\_tax\_rate_i * mileage_{i,y} * avg\_fuel_{i,y} \qquad (2)$$

The $mileage_{i,y}$ is computed based on the car stock and the average mileage of each car type in each year:

$$mileage_{i,y} = car\_stock_{i,y} * avg\_mileage_{i,y} \qquad (3)$$

In a similar approach as fuel tax estimation, the revenue from electricity tax equals the sum of the electricity taxes of all EV types that charge from the electricity grid, namely BEV and PHEV.

$$e\_tax_y = \sum_i e\_tax_{i,y} \qquad (4)$$

For each vehicle type, the tax is estimated as the multiplication of the tax rate per kWh consumed (€/kWh) and the total annual electricity consumption (which equals the multiplication of total annual mileage and average electricity consumption per kilometre).

$$e\_tax_{i,y} = e\_tax\_rate_i * mileage_{i,y} * avg\_e_{i,y} \qquad (5)$$

The data sources used for the computation is summarised in Appendix A.

*4.2. An overview of the multi-agent transport simulation (MATSim)*

MATSim is a software to implement large-scale multi-agent transport simulation that has been applied and developed by many scholars (Horni et al., 2016). It simulates transport users as individual agents conducting their travel activities simultaneously and interacting with each other. The physical environment in MATSim provides facility to simulate real world case studies and capture the traffic dynamics well. This is especially important for an analysis of traffic congestion. MATSim also allows accounting for individuals' heterogeneity, which is important to examine the distribution effects of policies. Each transport user has their own travel plans and a scoring function. Transport users can optimise their travel behaviours in an evolutionary iterative approach to maximise their scores.

Figure 4 illustrates the framework of MATSim with five main stages:

(i) Initial transportation demand generation: The initial demand of each agent often comes from a survey on the daily travel plans of a real population. Based on this information, MATSim generates a fixed number of daily plans for each agent. In this paper, the initial demand data come from the open source scenario of the Greater Berlin (Ziemke et al., 2019), which was based on the year 2008 (Gerd-Axel, 2009; Lenz et al., 2010).

(ii) Network loading with mobsim (mobility simulation): Mobsim loads the traffic network based on the provided data of road network and vehicles and allows agents to perform one of their daily plans and interact with other agents in the simulated physical environment. The plan selection depends on its recorded score. The data of the road network and vehicles are also taken from the work of Ziemke et al. (2019).

(iii) The scoring function is a sum of the utility related to the activity as well as the travel. All the plans that are executed during the simulations will be scored using this scoring function. Based on the recorded scores, agents can replan their travel activities to increase their utility. Our paper modifies the default scoring function so that it becomes income-dependent, which is important for the analysis of distribution effects (see section 4.3).

(iv) Replanning: In MATSim, agents can either choose among the fixed daily plans or explore new strategies through rescheduling activity, re-routing, or shifting mode. In our simulation, we allow agents to explore and innovate their travel plans through all available options in the first 80% of the iterations. For the last 20% of the simulation, the new strategies are switched off, and agents only chose among the fixed plans. This setting follows the previous work (Kaddoura and Nagel, 2019).

(v) Analyses: Stages (ii), (iii), and (iv) are in a loop for a pre-defined number of iterations. Afterwards, the outputs from the simulation are used for the analysis.

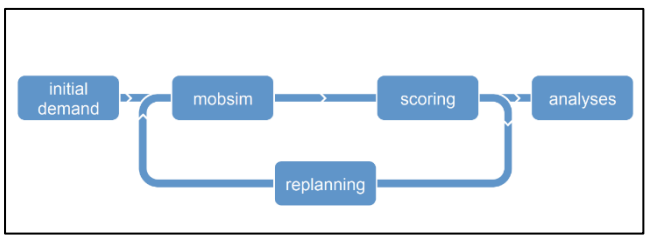

Figure 4: The MATSim framework. Source: matsim.org. The "initial demand" of each agent goes to "mobsim" (mobility simulation), which allows all agents to perform their plans and interact with each other. For each agent, different travel plans are generated and scored using a utility or "scoring" function. By comparing the scores of different plans, agents can "replan" their travel to achieve higher utility. The simulation runs automatically for the pre-defined number of iterations. Finally, the output from the simulation can be used for "analyses".

Depending on the research questions, different extensions or modules could be applied to MATSim to simulate different scenarios of traffic demand. In this paper, the module of "Decongestion" is applied. This allows a "toll controller agent" to be added to the simulation environment. This controller aims to correct the traffic delay and interacts with all transport users through the toll charging. More details on the congestion pricing method are provided in section 4.4. For further details of MATSim, readers are referred to its documentation (Horni et al., 2016).

*4.3. Income-dependent scoring function and conversion between utility and monetary terms*

In this section, we explain the modification of the standard scoring function in MATSim so that it can account for the income level of the individuals. This is similar to the approach applied in a previous study on road tolls in Switzerland (Kickhöfer et al., 2010).

The standard scoring function in MATSim is formulated as follow:

$$S_{plan} = \sum_{q=0}^{N-1} S_{act,q} + \sum_{q=0}^{N-1} S_{travel,mode(q)} \tag{6}$$

where $S_{act,q}$ includes the utility from performing the activity, waiting time, coming too late/early, and when the activity is too short[2] respectively ordered in the following equation:

$$S_{act,q} = S_{dur,q} + S_{wait,q} + S_{late.ar,q} + S_{early.dp,q} + S_{short.dur,q} \tag{7}$$

and $S_{travel,mode(q)}$ includes the travel-related cost such as constant cost of the mode, disutility of spending time on traffic, monetary costs (e.g., tolls), distance-related costs, and the transits of public transport:

$$S_{travel,mode(q)} = C_{mode(q)} + \beta_{trav,mode(q)} t_{trav,q} + \beta_m \Delta_{m,q} + (\beta_{d,mode(q)} \\ + \beta_m \gamma_{d,mode(q)}) d_{trav,q} + \beta_{transfer} x_{transfer,q} \tag{8}$$

We adjust the monetary costs $\beta_m \Delta_{m,q}$ in the equation of travel-related utility [equation (8)] to introduce the income-dependent feature. Instead of a constant global marginal utility of money $\beta_m$, an income-dependent agent specific marginal utility of money $\beta_{m,n}$ enters the scoring function:

$$\beta_{m,n} = \frac{\beta_m * population\ average\ income}{y_n} \tag{9}$$

where $y_n$ is the income of the agent $n$. By doing this, the agent-specific marginal utility of money is obtained by scaling the global marginal utility of money according to the agent's income relative to the population average income. This means that agents with lower incomes will have a higher marginal utility of money, reflecting the greater relative importance of money to them, while agents with higher incomes will have a lower marginal utility of money. In this study, as with other input data, agents' income profiles are based on the work of Ziemke et al. (2019).

Through the marginal utility of money, the utility and monetary terms are related in this study as follows:

$$utility_n = \beta_{m_n} * money\ amount_n \tag{10}$$

where $n$ is the agent index and $\beta_{m_n}$ is the agent specific marginal utility of money.

*4.4. Interval-based list congestion pricing*

---

[2] Penalties are introduced for the activities that are too short to encourage more realistic travel plans for agents. This expression is currently set at zero as there are no good data about this effect available.

The congestion pricing method in this paper is called interval-based list pricing (LP), which is based on the work of Kaddoura and Nagel (2019). This method sets the toll purely based on the traffic delay data on a specific road within a pre-defined time interval. Figure 5 illustrates this process.

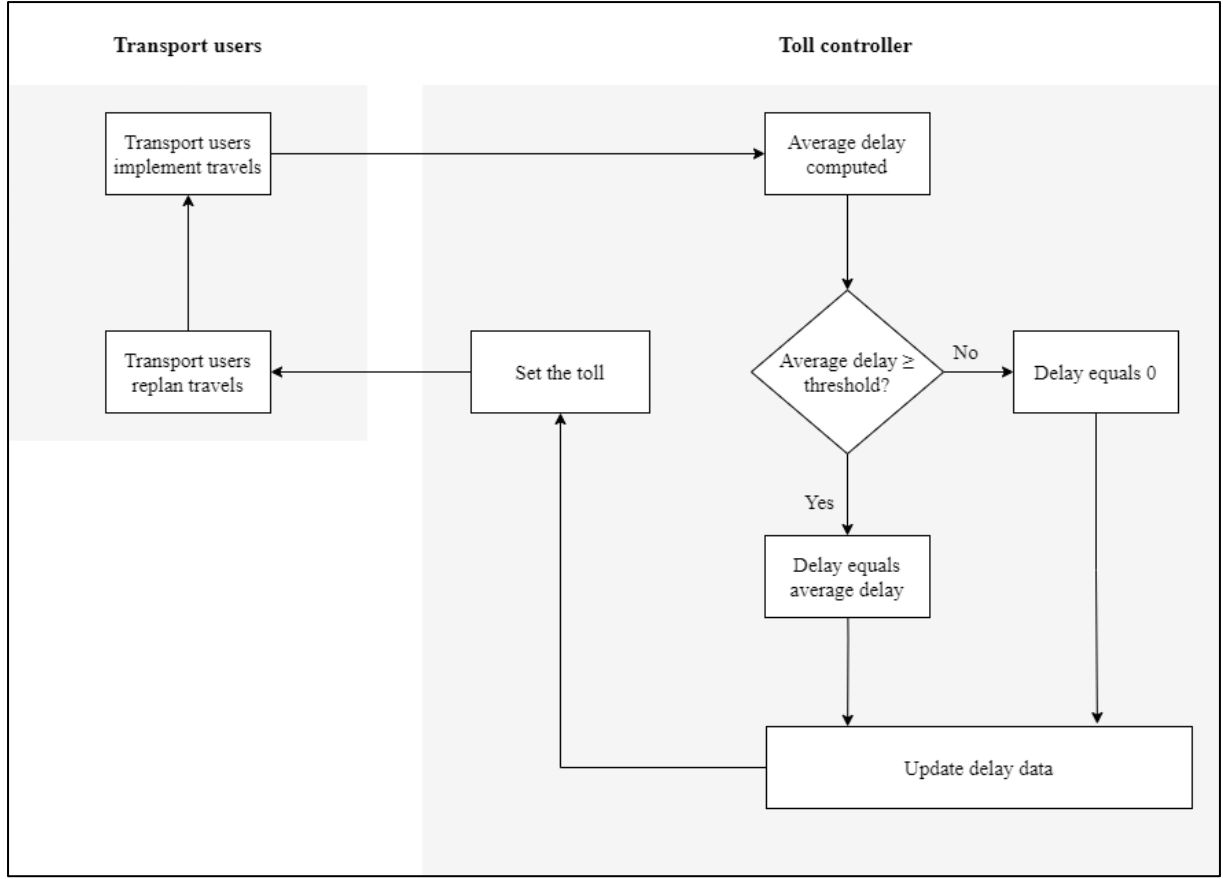

Figure 5: LP congestion pricing process in MATSim. In each simulation iteration, while transport users implement their travel plans, the traffic delay is computed for each road in a pre-defined time interval. Then the threshold of a minimum delay is used to determine the value of the delay to be updated in the data. Next, the tolls are set according to the delay level. Based on the new information (the toll), transport users replan their travel to maximise their utility. Then they implement their new travel plans in the next iteration and the loop continues.

In the first place, transport users implement their travel plans. At the same time, the average delay per road and time interval is computed as the difference between the actual travel time and free speed travel time, averaged by the number of transport users on the road at the time:

$$d_{r,t,k}^0 = \sum_{n=1}^{N_{r,t,k}} \frac{t_{r,t,n,k}^{act} - t_r^{free}}{N_{r,t,k}} \qquad (11)$$

where $r$ indicates the link, $t$ is the time interval, $k$ denotes the iteration, $n$ denotes the agent, $N_{r,t}$ is the total number of agents leaving link r in time interval t, $t^{act}_{r,t,n,k}$ and $t^{free}_r$ is the actual travel time and the free speed travel time, respectively.

Then, the average delay is compared with a specific threshold $d^{min}$. If the average delay is higher than this threshold, the delay value is updated to be equal to the average delay computed. Otherwise, delay is updated to be zero:

$$d_{r,t,k} = \begin{cases} d^0_{r,t,k} & for\ d^0_{r,t,k} \geq d^{min} \\ 0 & for\ d^0_{r,t,k} < d^{min} \end{cases} \quad (12)$$

Next, the toll setting is implemented according to the following equation:

$$m_{r,t,k} = \max\{0, K_p * d_{r,t,k}\} \quad (13)$$

where $m_{r,t,k}$ is the monetary payment fee and $K_p$ is the parameter pre-defined in the simulation.

As can be seen from Equation (13), the toll payment cannot be negative thanks to the "max" command, and it depends on the delay level of the road segment within a specific time and iteration ($d_{r,t,k}$). That is, the toll is computed automatically based on the delay data and there are no constraints on the total toll revenues. Higher delays lead to higher toll payments, which impacts travellers' utility (often negatively). Travellers now need to replan their travels to reduce the costs of the toll and maximise their own utility. Thanks to this interaction, the congestion toll can incentivize transport users to adjust their travel plans and reduce traffic congestion.

The parameter setting in this paper follows the previous work of Kaddoura and Nagel (2019).

*4.5. Upscale the simulation results*

The output from the simulation represents daily travel behaviours of 10% of the region's adult population during a representative day (see Tchervenkov et al. (2018) on upscaling from a representative day to an annual value). From the output, important indicators such as total toll revenue, social welfare, travel time and distance, etc., are computed. To present the results as annual values for the whole region, these numbers are upscaled linearly:

$$X^a{}_B = X^d{}_S * 10 * 365 \quad (14)$$

In Eq. (14), $X^a{}_B$ is the annual value for the greater Berlin area, $X^d{}_S$ denotes the daily value from the simulation.

5. **Results and discussion**

In this section, the simulation results for the greater Berlin area are discussed. First, an overview of the toll payments that are charged automatically based on the traffic delay is presented. Then, the effectiveness of congestion pricing is assessed through three key criteria: its role in substituting fuel tax in a world of EVs, its contribution to the society, and its distribution effects.

## 5.1. Congestion toll payments

Table 1 summarises the toll payments that transport users are charged in the last simulation iteration of three scenarios. Note that the traffic data of the case study is based on the year 2008. All the simulated monetary values have been adjusted for inflation using consumer price index (CPI, 2025 = 100) and presented as the 2025's values.

Table 1: Congestion toll payments and total toll revenue for the greater Berlin area. Monetary values are inflation adjusted by consumer price index to represent 2025's values.

| Indicator | Reference | Congestion | Congestion+ |
|---|---|---|---|
| **Average Toll/Trip [€]** | 0 | 0.32 | 0.26 |
| **Maximum Toll/Trip [€]** | 0 | 5.98 | 4.03 |
| **Average Toll/Km [€]** | 0 | 0.03 | 0.02 |
| **Annual Toll Revenue [million €]** | 0 | 784 | 615 |

As can be seen, there are no tolls in Reference and the values are slightly higher in Congestion than Congestion+ scenario. This is predictable as the improvement in the road capacity and cheaper public transport assumed in the Congestion+ scenario resulted in less congestion. The toll per car trip under Congestion scenario is €0.32 on average and €5.98 at the most. Under Congestion+ scenario, the range is narrower with an average of €0.26 per car trip and €4.03 to the maximum. This is in a similar range of road tolls that were observed in other studies. For example, the average congestion charge per car trip simulated for Munich varies between €0.8 and €3 (Kaddoura et al., 2017).

The toll rate per travelled kilometre is around €0.03, which is around one-third of the toll rate for infrastructure for trucks in Germany (Euro 6 emission class, 7.5–11.99 t) (toll-collect.de, 2023). This rate is also close to the real-world proposed toll in many countries. For example, according to the national road-pricing program in Netherlands, the tariff for cars was set at €0.03–0.067/km (Arnold et al., 2010). The approximation of the simulated toll rate and the world's currently applied policies indicates that the simulation results are realistic, and the follow-up assessment of the policy based on the simulation outputs can provide useful insights for politicians. Furthermore, the suggested toll level seems more easily accepted as the charge is in a similar range as existent systems worldwide.

An annual revenue of €784 million could be collected for the greater Berlin area in Congestion scenario. Under Congestion+ scenario, the toll amount is slightly lower at €615 million. We compare these figures with the expected reduction in energy taxes in the region in 2030 in the next sub-section.

*5.2. Congestion pricing substituting energy tax loss in a world of electric mobility*

To evaluate the potential of congestion pricing substituting energy tax loss due to EV diffusion, we first estimate the energy tax revenues in the two states of Berlin and Brandenburg in 2030 as detailed in section 4.1. We then compare these estimates to the congestion toll revenues generated from our simulation.

Figure 6 illustrates the estimated car stocks and the energy taxes therefrom for the Berlin-Brandenburg region. As can be seen, there will be a significant reduction in ICEs from roughly 2.45 million in 2023 to 1.6 million in 2030. The proportion of ICEs out of the total car stocks decreases from 91% to 63% during this period, increasingly replaced by EVs. This leads to a large reduction in the fuel tax revenue, which is projected to decrease by almost a haft from around €1.3 billion in 2023 to €733 million in 2030. At the same time, the increase in EVs also generates additional funds from electricity taxes. In 2030, electricity taxes from BEVs and PHEVs are estimated at roughly €41 million. On average, the tax rate per kilometre travelled is €0.03/km for fuel taxes and only €0.005/km for electricity taxes. Hence, although revenue from electricity taxes will increase with the growing number of EVs, this increase is insufficient to offset the substantial reduction in fuel tax revenues. As a result, the total energy tax revenue from cars, including both fuel and electricity taxes, will be remarkably lower than the average value during the last decade (2014–2023), as depicted in Figure 6. In 2030, there is a shortfall of around €671.6 million in Berlin-Brandenburg region to maintain the energy taxation revenue at its last decade's average value.

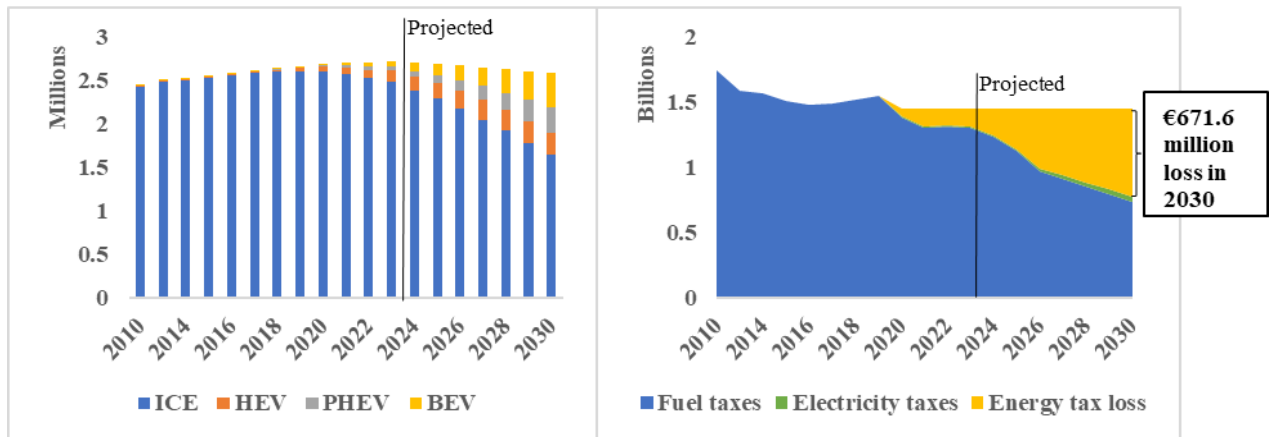

Figure 6: Berlin-Brandenburg region's car stocks and energy taxes (including fuel and electricity taxes) therefrom. Source: Own illustration. Monetary values are inflation adjusted based on CPI (2025 = 100). Real historical data are until 2023. The data for 2024-2030 are estimation under the scenario of dynamic transition to EVs according to the work of Betz et al. (2021). More details of the estimation method are presented in section 4.1.

Comparing with the net revenue reduction of € 671.6 million, we find that congestion tolls for the greater Berlin area alone can offset the two states' revenue loss. As shown in Figure 7, an annual revenue of around €784 million can be collected from congestion tolls in the Congestion scenario. This surpasses the energy tax loss of the region in 2030, generating a surplus for other public investments. Even when the traffic system is expected to improve and the congestion toll could be lower as assumed in the Congestion+ scenario, a revenue of €615 million from congestion pricing is roughly equivalent to the energy tax loss. Note that the congestion toll is computed based on the traffic demand data in 2008, when the traffic congestion was much lower than nowadays (The Local, 2023). Therefore, the actual congestion toll revenue in 2030 is expected to be higher. As can be seen, congestion pricing has a crucial contribution to the governmental revenue in both scenarios studied.

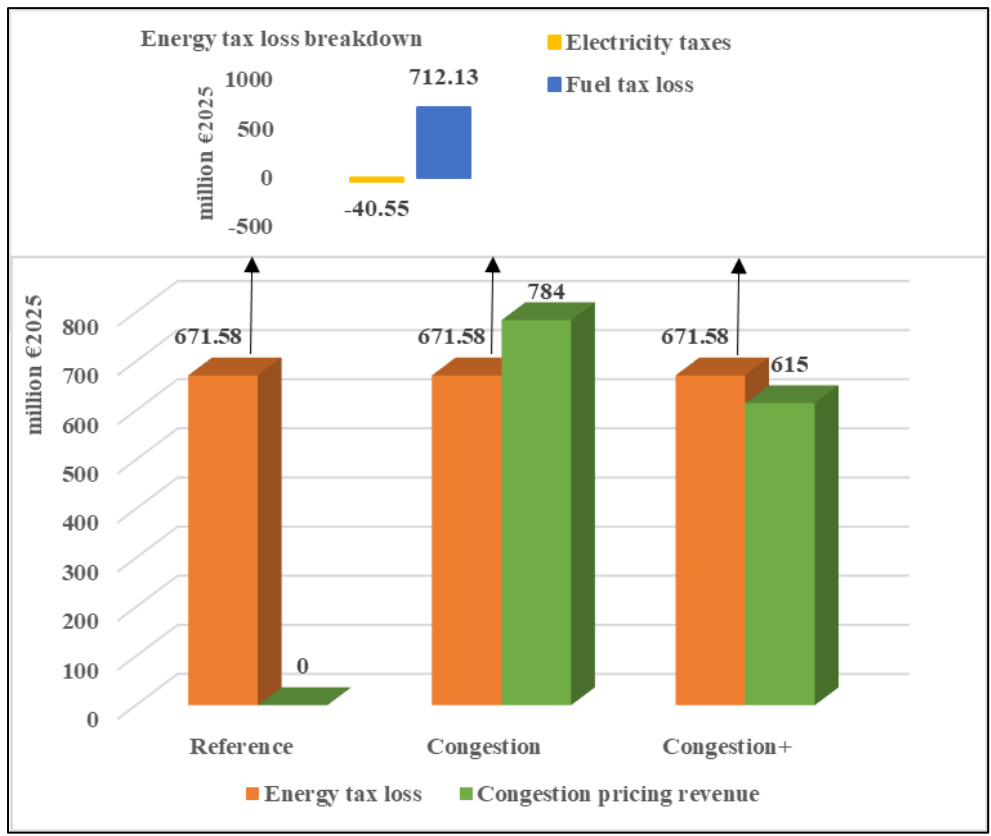

Figure 7: A comparison of energy tax loss for Berlin-Brandenburg states and congestion pricing revenues from the greater Berlin area. Source: Own illustration.

*5.3. Contribution to traffic management*

Table 2 summarises the changes to the traffic operation thanks to congestion pricing for the simulated area. The Congestion and Congestion+ scenarios are compared with the Reference scenario. Details of the simulation outputs for three scenarios are provided in Appendix B.

Table 2: Traffic changes resulting from congestion pricing in the greater Berlin area (compared to reference scenario)

|  | **Congestion** | **Congestion+** |
|---|---|---|
| **Total car delay change [million hours]** | -17.30 (-28.08%) | -27.40 (-44.47%) |
| **Car trips change [million]** | -69.44 (-2.82%) | -67.16 (-2.72%) |
| **Car travel distance change [million km]** | -608.35 (-1.93%) | -619.74 (-1.97%) |
| **Car shifted to public transport [million km]** | 510.91 | 576.09 |
| **Travel with departure time shift [billion km]** | 19.37 (45%) | 21.46 (50%) |

The table shows that the introduction of congestion pricing reduces the traffic delay in the region by over 28%. When congestion pricing is coupled with expanded road capacity and subsidies for PT, the reduction in traffic congestion can be up to 45%. The relief of traffic congestion mostly comes from a general reduction in car travel as well as an adjustments in travel routes and departure time to avoid the toll. Nearly 70 million car trips (~3%) are avoided annually when the congestion toll is introduced, which corresponds to over 608 million kilometres of less car travel (~2%). Not surprisingly, the car travel reduction is slightly lower in Congestion+ scenario because the additional road capacity makes congestion less likely and cheaper to resolve. A stronger shift from cars to PT is also observed in Congestion+ scenario, where PT is much cheaper. Around 580 million kilometres of car travel are shifted to PT in Congestion+ scenario, compared to 510 million in Congestion scenario.

A change in departure time is also observed frequently. Still on Table 2, around half of all travel trips change the departure time. Figure 8 provides additional information with respect to the mode shift. It shows that around 60% of those keeping driving cars (car2car) or changing from PT to cars (pt2car) are adjusting their travel time. Time rescheduling is much less relevant for those staying with PT (pt2pt), who are not charged by congestion toll, with over 80% of the trips keeping their departure time. Apparently, congestion toll seems to influence car users' travel plans and motivates them to avoid traffic jam. This contributes to reducing the overall traffic congestion.

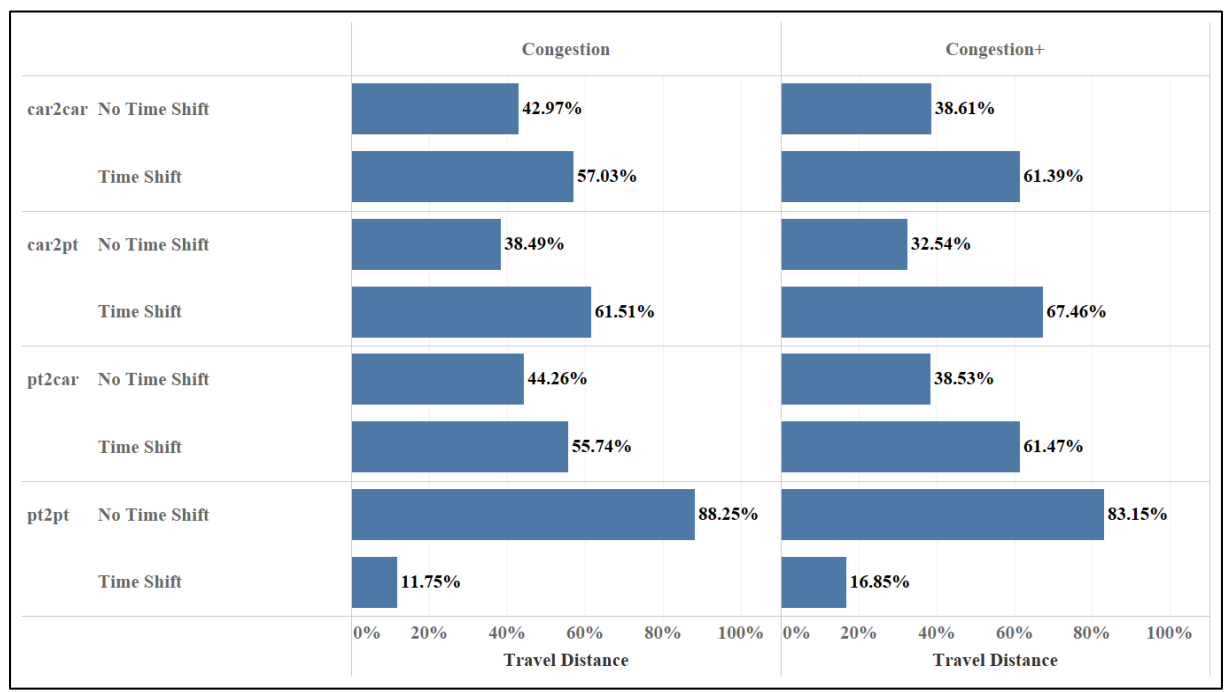

Figure 8: Travel time adaptation by mode change. Source: Own illustration. This figure shows the proportion of the travel that changed the departure time when congestion pricing was introduced, grouped by the mode change. Car2car = users keeping using cars; car2pt = users shifting from cars to public transport; pt2car = users shifting from public transport to cars; pt2pt = users keeping using public transport.

*5.4. Contribution to environment*

Congestion pricing can also contribute to reducing air pollution. We provide a rough estimation of $CO_2$ emission reduction resulting from reduced traffic delays and car travels, as illustrated in Figure 9.

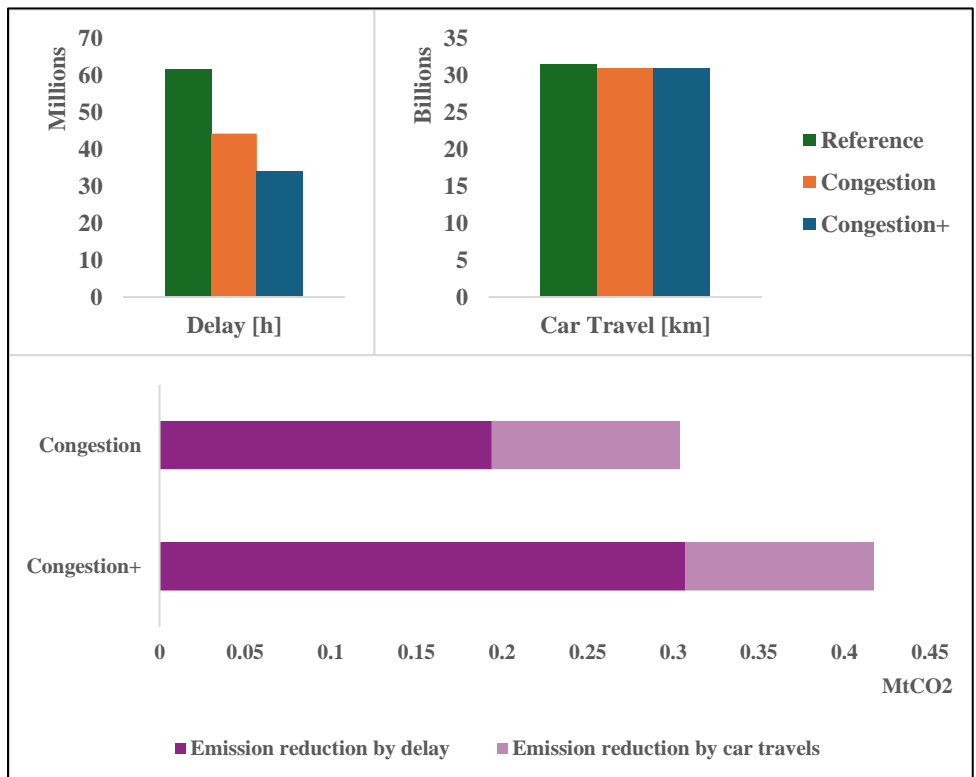

Figure 9: Congestion pricing reduces emissions via less traffic delay and car travels.

Compared to the Reference scenario, 17.3 (Congestion scenario) and 27.4 million hours of traffic delay (Congestion+ scenario) are avoided. Based on the estimation of annual emissions from the passenger car sector in Berlin and Brandenburg states (Kaddoura et al., 2022b) and the average daily travel time in Germany (BMVI, 2019), we calculate the emission rate per hour in studied region. Using this rate, we compute the emission reduction thanks to less traffic delay time. We assume that each hour of traffic delay emits equally to the average hourly emission rate, although the actual emissions during congestion time is typically higher than average (Choudhary and Gokhale, 2016). We show that the traffic delay reduction lowers the emissions in the region by 0.2–0.3 $MtCO_2$ per year. Furthermore, using the current emission factor of passenger cars in Germany (181.887 gram per vehicle-kilometre, average emission category and fuel type based on HBEFA 4.1, Germany, year 2020) (Kaddoura et al., 2022a), we also estimate the emissions that could be cut thanks to less car travels. The car travelling decrease of around 600 million kilometres corresponds to 0.11 $MtCO_2$ reduction per year. In total, there is a decline of 0.31 $MtCO_2$ (~5.3% of emissions in Reference scenario) thanks to congestion pricing. Together with larger road capacity and cheaper PT, the emission decrease is even stronger at 7.3%. As mentioned, our estimation neglects other mechanisms such as less congestion leading to less energy waste (Cramton et al., 2018) and thus less emissions. Consequently, the real impacts of congestion toll to the environment can be higher than our estimation.

*5.5. Social welfare and distribution effects*

This section examines the effects of congestion pricing on social welfare and its distribution effects across different income groups (referred to here as vertical distribution) and various areas within the metropolitan region (horizontal distribution).

The simulation results indicate that congestion pricing generates a significant social welfare surplus. This surplus arises as the total toll revenues exceed the disutility experienced by the population from paying the tolls. On average, each traveller's utility decreases by €0.11 per day when congestion fee is introduced, which totals approximately €191 million annually for the population of the simulated region. This figure is notably lower than the congestion toll revenue of €784 million (see section 5.1). This revenue not only offsets the utility loss but also generate a net social welfare gain of €593 million. The following sub-sections provide a detailed analysis of the welfare impacts, focusing on distributional effects.

*5.5.1. Impacts on social welfare by income group*

Vertical distribution effects often focus on the impacts of a policy on different income groups. Figure 10 illustrates this through the toll payment and utility changes. The highest income group (IG 1) comprises the top 10% of the earners in the sample, while the lowest income decile is IG 10, at the bottom of the chart. As can be seen from panel a, the payments are linearly positively correlated with the income level, where the very high-income pay the most. The highest income group pays an average daily fee of €4.3 under Congestion scenario and €3.4 under Congestion+ scenario. This is over 6 times higher than the charge for the lowest income group. In total, the highest income population contribute to 17% of the total toll revenue, compared to 2.5% from the lowest income. Consequently, under this congestion pricing policy, there is a high possibility of redistributing toll revenue from the high- to low-income residents.

Panel b of the figure shows the toll payments in utility unit, which are derived from the monetary payment values using marginal utility of money (see equation (10)). In this representation, the disparity in payments among income groups is considerably less pronounced (than the monetary payments). The toll costs range from 1.1 to 1.7 utility units for Congestion scenario and 0.8 to 1.4 utility units for Congestion+ scenario. This indicates that the toll was optimized such that the cost is evenly distributed across the entire population. Furthermore, this panel also shows that the toll seems the most expensive for middle income groups instead of the highest income group as suggested in panel a. One explanation for this comes from the elasticity in the travel demand and the marginal utility of money. Previous studies showed that gains from congestion pricing are small for those with inelastic or less elastic travel demand (Eliasson and Mattsson, 2006). The middle-low-income groups are often less flexible in their travel plans, especially the work-related travel (e.g., time to work, routes, …). They often have their workplace in the inner city making it difficult to avoid the charges. Furthermore, they also have higher marginal utility of money than high-income groups, which leads to more utility loss per monetary unit paid. Conversely, the lowest income decile in IG 10 pays the least both in terms of monetary values and utility units. Apparently, this income group includes a large proportion of the unemployed and students, whose travel demand is also much more elastic than the other income groups. Furthermore, since they also tend to use public transport rather than cars, they might be less affected by the toll and gain more if the toll revenue is spent on subsidising and/or improving PT.

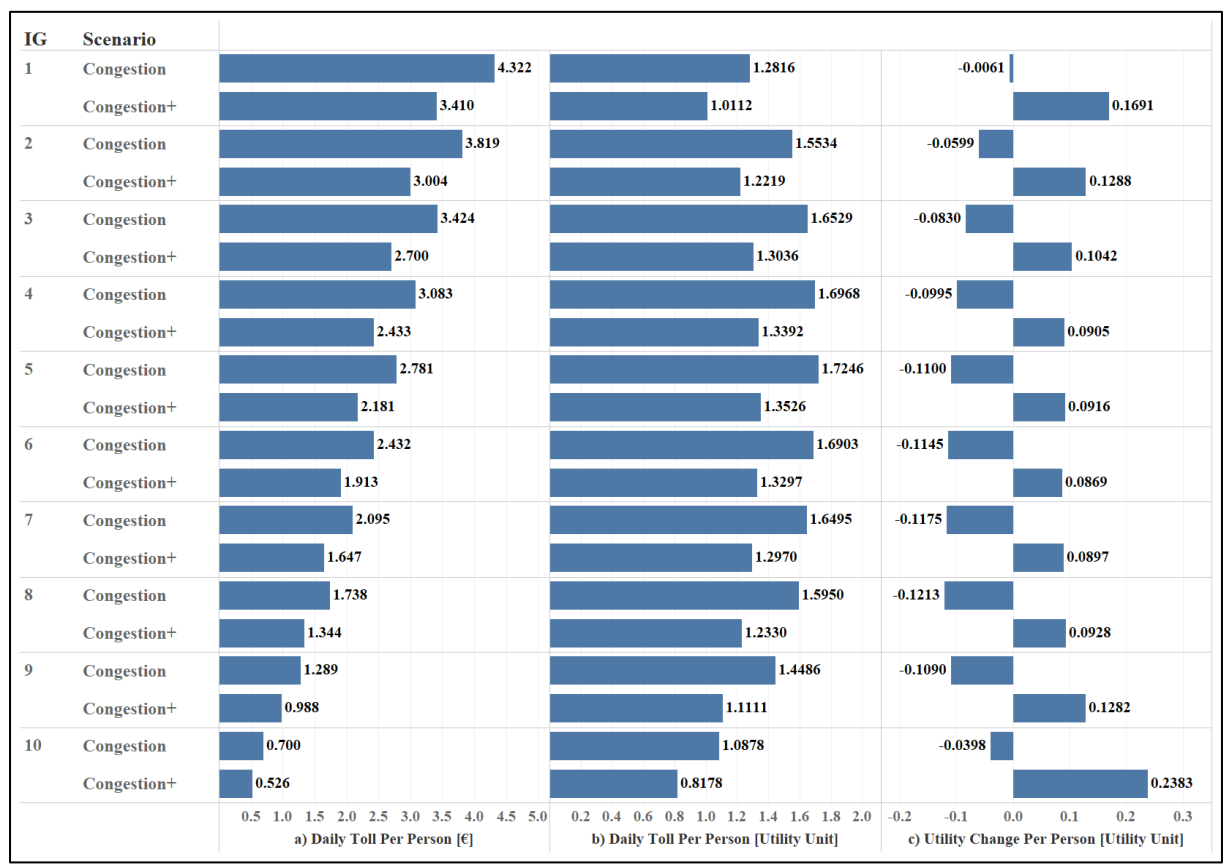

Figure 10: Daily toll payment and utility change of car users by income deciles. Source: Own illustration. The first decile is the highest income group. IG. = Income group.

Panel c of the figure shows the change in the average utility of each income group. In the Congestion scenario, the highest and lowest income deciles experience minimal utility loss, while the middle-low-income are more affected by the toll. There is a utility loss of around 0.12 unit per person for the IGs 6–8, compared to almost zero loss of utility for the IG 1 and around 0.04 utility unit for IG 10. In Congestion+ scenario, utility gains are observed for all income groups. The largest gain is for the bottom income group with an increase of utility by 0.24 unit per person. Following is the top income interval with 0.17. While the former group gains more from cheaper PT, the later mostly benefit from the larger road capacity and flow improvement, which makes car travelling more convenient. The utility increase is slightly smaller for middle-low-income groups. This indicates that the middle-low-income groups not only pay the most for the tolls (in terms of utility units) and are the worst affected population, but they also gain less from expanding road capacity or subsidising PT. Therefore, the highest compensation should come to the middle-low-income deciles, either through lump-sums or improving alternatives to cars.

*5.5.2. Impacts on social welfare by area*

Horizontal distribution effects are examined through the spatial plots in which the utility of residents is analysed based on home locations rather than income intervals. Figure 11 illustrates the toll payment and utility change when congestion pricing is introduced for Berlin area by geopolitical zones. Panel a describes the population distribution among the zones and shows that most of the population reside far from the city centre to the north-west, west, and south parts (zones 1, 2, 3, 8, 9, 11, 12). These regions are also among the most contributors to the total toll revenue and show the largest total utility loss, as can be seen from panel b and c. One might envision additional investments to improve the connections between these regions and the centre of Berlin to compensate for the utility loss.

If the average toll per resident (rather than the total toll of the area) is considered, as depicted in panel d, the highest toll rate per car trip is observed for the zones that are furthest from the city central area (zones 2 and 17). This explains the larger utility loss for those living further from the city central area (as can be seen from panel e of the figure), especially for the residents in the north-and south-eastern zones. This slightly deviates from the observations in panels b and c, when the total values are considered. This highlights the importance of indicator selection in assessing the impacts of congestion pricing. As illustrated, assessments based on the aggregate utility loss might favour the region with large population. As a result, less populous regions might be under-invested, and the utility loss of the residents here might be overlooked. Therefore, both aggregate and average values should be accounted for.

The results for Congestion+ scenarios are mostly identical and are presented in Appendix C.

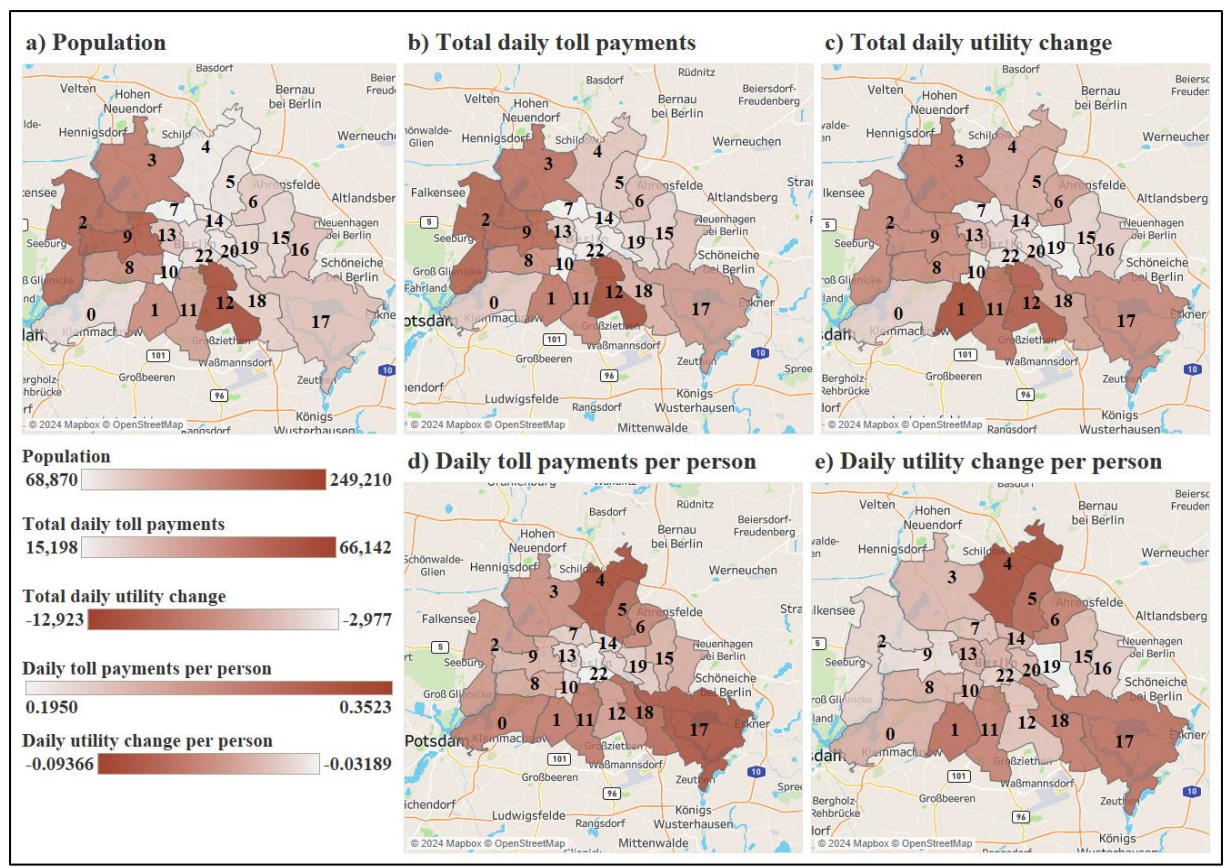

Figure 11: Daily toll payment [€] and utility change by geopolitical zones in Berlin under Congestion scenario. The zones are numbered 0–22. Inner zones include the central and adjacent zones (7, 10, 13, 19, 20, 21, 22). The others are outer zones.

## 6. Conclusions

The transition to electric mobility will decrease energy tax revenues worldwide substantially. Our estimates for Germany show that the tax revenue per kilometre driven decreases by a factor of six from €0.03/km for internal combustion engine cars to €0.005/km for electric vehicles. In 2030, the fuel tax revenue in Germany is estimated to be reduced by €12.5 billion compared to 2020's values. At the same time, traffic congestion remains a persistent global problem and congestion pricing is well known as an effective solution for this – but rarely implemented in practice. Taken together, we started this paper with the hypothesis that the emerging shortfall in energy tax revenue presents a timely opportunity to seriously discuss and implement intelligent road pricing. The benefit would be two-fold: First, governments could compensate tax reductions without discouraging the desired shift to electric mobility.

Second, tolls are a welfare-increasing way to generate revenues because they improve traffic flows and correct externalities.

This paper thus examines the effectiveness of congestion pricing as a sustainable alternative revenue stream for fuel taxes, which are expected to decrease substantially due to the diffusion of electric vehicles. To that end, a real-world case study of the greater Berlin area in Germany, including the adult population residing in Berlin and the commuters from Brandenburg to Berlin, is analysed. We estimate the total energy tax loss of the region in 2030, assuming a dynamic growth of electric vehicles, and compare it with the revenue from congestion tolls. Furthermore, the impacts of congestion toll on the traffic management, environment, as well as the social welfare and distribution effects are also quantified. Congestion pricing is simulated using multi-agent traffic simulation software MATSim. Three scenarios are analysed: Reference scenario — traffic demand without any tolls; Congestion scenario — keeping the setting of Reference scenario unchanged, except congestion pricing introduced; Congestion+ scenario — congestion pricing introduced and extended road capacity and subsidised public transport.

The analysis shows that congestion pricing contributes significantly to the national revenue. The toll revenues (€784 million in Congestion scenario and €615 million in Congestion+ scenario) can effectively fill in for the energy tax loss of €671.6 million due to electric vehicle diffusion in the simulated region. This is achieved despite the tolls being set solely based on traffic patterns, without any constraints on the total revenue amount.

The introduction of congestion tolls also successfully incentivises transport users to adjust their travel behaviour, which reduces traffic congestion and related negative externalities substantially. Around 3% of car trips are either dropped or changed to public transport and around 60% of car travellers shift their departure time to avoid rush hours. As a result, there is a decrease in the traffic delay time by 28–45%. Noteworthily, the traffic delay and car travel reduction lead to a decline in $CO_2$ emissions by 5–7%. The transport infrastructure is also used more efficiently, which reduces the pressure and long-term depreciation of the roads. Therefore, in addition to being an important source of funds for the governmental revenue, congestion pricing is also an appropriate measure to reduce externalities and enhance the efficiency in the road transport sector.

Regarding distribution effects, first, the analysis by the population's income intervals or vertical distribution effects show that congestion pricing could be progressive, i.e. travellers with higher income pay more than those with lower income. The highest income group pays over 6 times higher than the lowest income group, contributing to 17% of the total toll revenue, compared to 2.5% from the lowest income. The investments in road infrastructure and subsidies for public transport are the most beneficial to the lowest income interval. Middle-low-income groups have the largest decline in utility, which implies a redistribution focus on this population segment. Second, the impacts of congestion pricing on residents are varied and cannot be captured completely by vertical distribution effects. Horizontal distribution effects play a role in explaining the utility changes of residents and should be considered in formulating the revenue redistribution plan. There is an overall tendency of higher utility loss for residents living far from the city central areas.

Note that the congestion toll revenues for the simulated region amount to €784 million, which significantly exceeds the population's utility loss of €191 million. This surplus not only covers the energy tax loss but also allows for redistribution. Redistribution efforts should prioritize middle- and low-income groups, as well as outer boroughs. For example, lump-sum payments could compensate for the utility loss. At the same time, a transparent investment plan of the toll revenue to upgrade the transport systems connecting the outer zones and the city centre should be presented. Finally, we show that priority for redistribution and investments should be given to densely populated boroughs as well as regions of high average congestion costs (but might be less populous).

The insights from this study have global importance, considering the transitions to EVs occurring worldwide. The revenue loss could be especially problematic for the countries with already high market share for EV sales such as China, the US, and other European countries such as Norway, Iceland, Sweden, and the Netherlands or those rather dependent on energy taxes such as Honduras, Bulgaria, Croatia, Latvia, and Pakistan. We show that intelligent road tolls such as congestion pricing could be an effective solution for this problem. Furthermore, the policy assessment method applied in this study can also be replicated for other case studies and is thus an important tool for politicians, local authorities, researchers, and other stakeholders.

Last but not least, we recommend additional analyses in the future that either require additional data or are beyond the scope of this paper. For example, the transaction costs of implementing and monitoring the toll system should be included in the analysis. In addition, other impacts of congestion charge to the land use, e.g., parking pressures around the charged area, residents moving their houses to avoid the toll, etc., are also important and should be analysed as well.

**Acknowledgements**

FMü gratefully acknowledge the financial support by the Federal Ministry of Education and Research, project number 19FS2032C as well as the German Federal Government, the Federal Ministry of Education and Research, and the State of Brandenburg within the framework of the joint project EIZ: Energy Innovation Center, project numbers 85056897 and 03SF0693A, with funds from the Structural Development Act (Strukturstärkungsgesetz) for coal-mining regions.

TNN gratefully acknowledge financial support from the Federal Ministry of Education and Research of Germany in the Ariadne 2 project (03SFK5O0-2).

**Appendix A.    Data sources**

| Data | Source |
|---|---|
| Avg. electricity consumption for BEVs | Energy consumption of full electric vehicles<br>https://ev-database.org/cheatsheet/energy-consumption-electric-car |
| Avg. electricity consumption for (P)HEVs | Fuel economy guide – Model 2023<br>https://www.fueleconomy.gov/feg/pdfs/guides/feg2023.pdf |
| Avg. fuel consumption for ICE-petrol and ICE-diesel [2004-2021] | Verkehr in Zahlen 2022/2023<br>https://bmdv.bund.de/SharedDocs/DE/Publikationen/G/verkehr-in-zahlen-2022-2023-pdf.pdf?__blob=publicationFile |
| Avg. fuel consumption for HEV-petrol, HEV-diesel, and PHEV-petrol [2006-2019] | Fuel Economy in Germany<br>https://www.iea.org/articles/fuel-economy-in-germany |
| Avg. mileage by vehicle type and power train [2014-2021] | Verkehr in Kilometern (VK), Zeitreihe 2014-2021<br>https://www.kba.de |
| Car stock by vehicle type and power train [1960-2022] | Bestand an Kraftfahrzeugen und Kraftfahrzeuganhängern nach Bundesländern, Fahrzeugklassen und ausgewählten Merkmalen<br>https://www.kba.de |
| Tax rate for petrol and diesel [2021] | Benzinpreise und Dieselpreise - aktuelle Preise, News, Trends und Prognosen<br>https://www.tanke-guenstig.de/Benzinpreise |
| Tax rate for electricity | BDEW-Strompreisanalyse Jahresbeginn 2023<br>https://www.bdew.de/service/daten-und-grafiken/bdew-strompreisanalyse/ |

Appendix B.  Key simulation results

We present the key simulation results as raw data in Table B.1. The raw data are then upscaled to the full population level and reported as annual values, as summarised in Table B.2.

Table B.1: Raw data for key simulation results, which are the daily values for 10% of the population in the greater Berlin region.

| Indicator | Reference_raw | Congestion_raw | Congestion+_raw |
| --- | --- | --- | --- |
| **Average toll/trip [€]** | | 0.32 | 0.26 |
| **Maximum toll/trip [€]** | | 5.98 | 4.03 |
| **Average toll/km [€]** | | 0.03 | 0.02 |
| **Daily toll revenue [€]** | | 151891.75 | 119146.85 |
| **Total daily car delay [hours]** | 16878 | 12138 | 9372 |
| **Total daily car trips** | 675834 | 656808 | 657433 |
| **Total daily car travel [km]** | 8638877.63 | 8472205.74 | 8469086.78 |
| **Average distance per car trip [km]** | 12.78 | 12.90 | 12.88 |

Table B.2: Key simulation results reported in the paper. Aggregate data are upscaled from the raw data by 10 to get the data for the full population level of the region and then upscaled by 365 to get the annual values. Monetary data are inflation adjusted using CPI (2025=100).

| Indicator | Reference | Congestion | Congestion+ |
| --- | --- | --- | --- |
| **Average toll/trip [€]** | | 0.32 | 0.26 |
| **Maximum toll/trip [€]** | | 5.98 | 4.03 |
| **Average toll/km [€]** | | 0.03 | 0.02 |
| **Annual toll revenue [million €]** | | 784.26 | 615.19 |
| **Annual car delay [million hours]** | 61.60 | 44.30 | 34.21 |
| **Annual car trips [billion trips]** | 2.47 | 2.40 | 2.40 |
| **Annual car travel [billion km]** | 31.53 | 30.92 | 30.91 |
| **Average distance per car trip [km]** | 12.78 | 12.90 | 12.88 |

Appendix C.     Horizontal distribution effect analysis for Congestion+ scenario

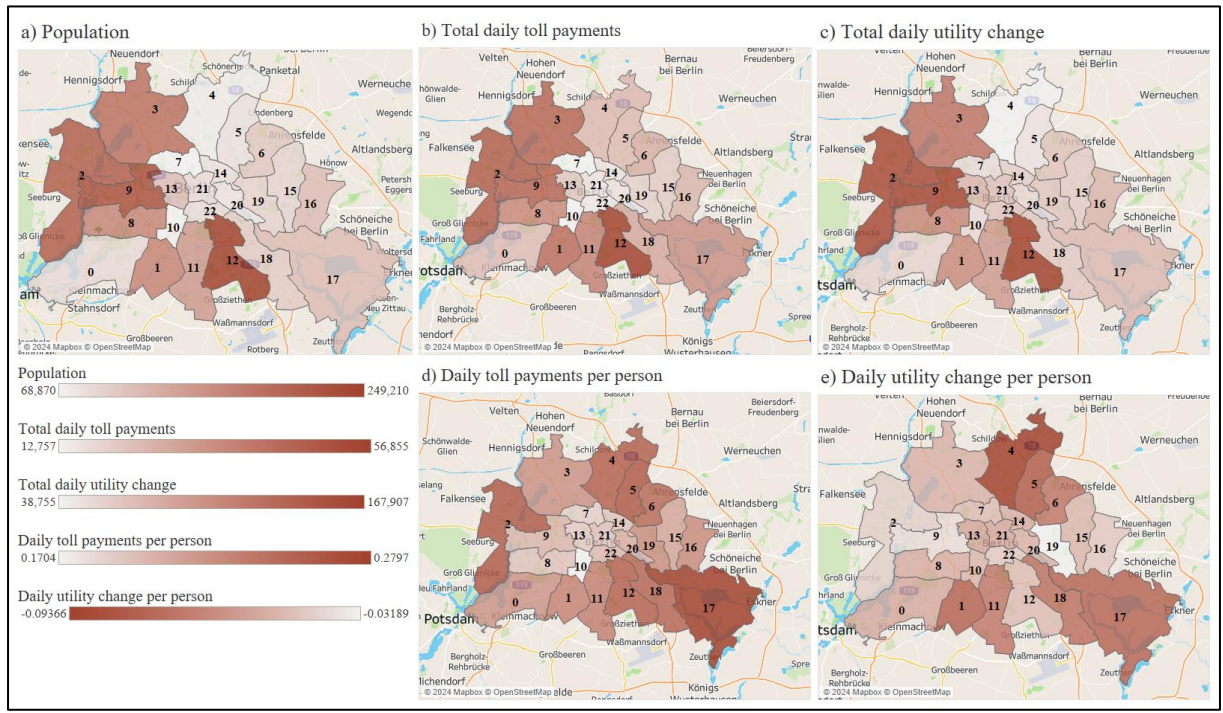

Figure C.1: Daily toll payment [€] and utility change [€] by geopolitical zones in Berlin under Congestion+ scenario. The zones are numbered 0–22. Inner zones include the central and adjacent zones (7, 10, 13, 19, 20, 21, 22). The others are outer zones.